\documentclass[aps,prc,reprint,superscriptaddress,nofootinbib]{revtex4-1}
\usepackage{graphicx}
\usepackage{multirow}
\usepackage{comment}
\usepackage{color}

\usepackage{amssymb}
\usepackage{amsmath}
\usepackage{color}
\usepackage{lineno}

\newcommand{\zbx}{Z_x(\vec{k}_b,\xi,\vecr_x)}
\newcommand{\psix}{\psi_x^{0}(\vec{k}_b,\vecr_x)}

\renewcommand{\Im}{\mathop{\mathrm{Im}}}

\def\nuc#1#2{\relax\ifmmode{}^{#1}{\protect\text{#2}}\else${}^{#1}$#2\fi}

\newcommand{\vecr}{{\vec r}}

\newcommand{\be}{\begin{eqnarray}}
\newcommand{\ee}{\end{eqnarray}}

\newcommand{\bwt}{\begin{widetext}}
\newcommand{\ewt}{\end{widetext}}

\begin{document}

\title{Reexamining closed-form formulae for inclusive breakup: 
Application to deuteron and $^6$Li induced reactions}


\author{Jin Lei}
\email[]{jinlei@us.es}

\affiliation{Departamento de FAMN, Universidad de Sevilla, 
Apartado 1065, 41080 Sevilla, Spain.}

\author{A. M. Moro}
\email[]{moro@us.es}

\affiliation{Departamento de FAMN, Universidad de Sevilla, 
Apartado 1065, 41080 Sevilla, Spain.}


\begin{abstract}
The problem of the calculation of inclusive breakup cross sections in 
nuclear reactions is reexamined. For that purpose, the post-form theory 
proposed by  Ichimura, Austern and Vincent [Phys. Rev. C32, 431 (1985)] 
is revisited,  and an alternative derivation of the non-elastic breakup 
part of the inclusive breakup is presented, making use of the 
coupled-channels optical theorem. Using the DWBA version of this model, 
several applications to deuteron and $^6$Li reactions are presented and 
compared with available data. The validity of the zero-range approximation 
of the DWBA formula is also investigated by comparing zero-range  with 
full finite-range calculations. 
\end{abstract}


\pacs{24.10.Eq, 25.70.Mn, 25.45.-z}
\date{\today}%
\maketitle

\section{Introduction \label{sec:intro}}

The breakup of a nucleus into two or more fragments is an 
important mechanism occurring in nuclear collisions, particularly 
when one of the colliding nuclei is weakly bound. The analysis of this 
kind of processes has provided useful information on the structure of the 
broken nucleus, such as binding energies, spectroscopic factors  and 
angular momentum (e.g.~\cite{Nak09,Nak12}), and has contributed to the 
understanding of the dynamics of the reactions among composite systems. 


In the simplest scenario, in which the projectile is broken up into two 
fragments, these processes can be schematically represented as 
$a+A \rightarrow b+ x+ A$, where $a=b+x$. From the theoretical point of 
view, this problem is difficult to treat because one has to deal with 
three-body final states. When the state of the three outgoing fragments ($b$, $x$ and $A$) is fully determined,  the reaction is said to be {\it exclusive}. If, in addition, the three 
particles are emitted in their ground state,  the corresponding cross 
section is referred to as {\it elastic breakup} (EBU). In this case,  
the reaction can be treated as an effective three-body problem interacting 
via some effective two-body interactions. Although the rigorous formal 
solution of this problem is given by the Faddeev formalism 
\cite{Fad60,Gloc83}, the difficulty of solving these equations has led 
to the development of simpler approaches, such as the distorted-wave Born 
approximation (DWBA) \cite{Shy83}, the continuum-discretized coupled-channels 
(CDCC) method  \cite{Aus87} and a variety of semiclassical approaches 
\cite{Typ94,Esb96,Kid94,Cap04}. 

 A more complicated situation occurs when the final state of one or more fragments  
is not specified. In this case, the reaction is said to be  {\it inclusive} with respect to this unobserved particle(s). 
This is the case of reactions of the form 
 $a+A \rightarrow b + B^*$, where $B^*$ is 
any possible configuration of the $x+A$ system. This includes the breakup processes in which $x$ is elastically scattered by $A$, which corresponds to the EBU defined above, 
but also  breakup accompanied by target excitation, particle(s) exchange between $x$ and $A$, $x$ transfer to  $A$,  and the fusion of $x$ by $A$, which are globally referred to as 
{\it non-elastic breakup} (NEB).   The total inclusive breakup (TBU) will be therefore the sum of EBU and NEB components, i.e. TBU=EBU+NEB.  
Measured observables usually correspond 
to single or double differential cross sections with respect to the angle 
and/or energy  of $b$  and hence include both EBU and NEB contributions.

The evaluation of NEB cross sections are needed, for example,  in the 
calculation of total fusion cross sections in reactions induced by 
weakly-bound projectiles (e.g.~$^6$Li, $^7$Li, $^{9}$Be). A significant 
fraction of the total fusion cross section comes from incomplete fusion 
(ICF), in which only part of the projectile fuses with the target, the 
other fragment surviving after the collision \cite{Das02}.  Although many 
theoretical efforts have been made to develop suitable models to 
calculated ICF cross sections \cite{Can98,Hus04,Has09}, the unambiguous 
calculation of CF and ICF within a fully quantum mechanical model remains 
a challenging problem \cite{Bose14,Tho04}. Because the ICF is part of the 
inclusive breakup, the study of inclusive breakup reactions may lead in 
turn to a better understanding of ICF. 

A related problem is that of the indirect determination of  neutron-induced cross sections on short-lived nuclei, from a surrogate reaction 
which gives rise to the same compound nucleus \cite{Esc12}. This is the case, for example, of the process $A(n,f)$ (where $f$ is a fission fragment) for which the surrogate reaction $A(d,pf)$ may be used. To extract the cross section for the former, one needs to know the fraction of protons produced in 
the surrogate reaction which are accompanied by the formation of a $n+A$ compound nucleus. Therefore, the applicability
of the method requires the separation of the EBU component 
(which does not lead to compound-nucleus formation) from the NEB (which contains the absorption cross section). 


The calculation of inclusive breakup observables is  more involved than that for 
the exclusive ones because they require the inclusion of all the possible 
processes through which the particle $x$ can interact with the target $A$. 
Given the large number of accessible states, this procedure is 
unpractical in most cases. As an alternative to this approach, one may 
try to replace the physical final states by a set of representative states
(also named {\it doorway} states).  These can  be taken, for example, as the eigenstates of the $x+A$ Hamiltonian in a mean-field potential.
As long as the basis used to describe these 
final states is complete, one may argue that the sum over these 
representative states should provide results close to those obtained if 
the sum were done over the true physical states. This procedure, referred 
to in some  works as {\it transfer to the continuum} method, has 
been used recently with  rather success to describe some inclusive breakup 
reactions of weakly-bound projectiles at Coulomb barrier energies, 
such as $^{208}$Pb($^{8}$Li,$^7$Li) \cite{Mor03}, $^{208}$Pb($^6$He,$\alpha$) 
\cite{Esc07}, and $^{120}$Sn($^6$He,$\alpha$) \cite{Far10}. 
 However, despite this relative success, this method is based on a 
 heuristic approach rather than on a rigorous formal theory. Lacking 
 this formal justification, it is not clear how these {\it doorway} states 
 should be chosen and how the final calculated cross sections depend on 
 this choice. Another drawback of this approach is that it does not allow 
 to separate the contributions coming from EBU and NEB.

At intermediate energies (above $\sim$100~MeV/u), the problem can be 
greatly simplified using the {\it adiabatic} (fast collision) and 
{\it eikonal} (forward scattering) approximations,
which allows to 
obtain closed formulas of the inclusive process in terms of the absorption and survival 
probabilities of the unobserved particle as a function of the impact parameter. 
This approach has been used extensively in the analysis of nucleon removal 
(knockout) experiments at intermediate energies, in which typically the 
removed particle is not observed and only the momentum distributions of 
the residual core is measured (see, e.g.~Refs.~\cite{Tos01,Han03} and 
references therein).  These models, however, cannot be applied to low 
incident energies (a few MeV/u) and when the energy/momentum transfer
is large.


The problem of the calculation of inclusive breakup cross sections is 
nevertheless not new.   This problem was studied in detail by some 
groups since the late seventies, and several theories were proposed and applied. 
The aim of these theories was to derive closed-form formulas, in which 
the sum over final states of the $x+A$ system is formally 
reduced to some expectation value of the imaginary part of the $x+A$ 
optical potential.  In the pioneering works by Baur and co-workers 
\cite{Bud78,Bau80,Shy80}, the sum is done making use of unitarity and a 
surface approximation of the form factors of excited states of the 
residual nucleus.  These two approximations were avoided in later works 
by Udagawa and Tamura \cite{Uda81,Uda84}, who used a prior-form DWBA 
formalism, and by Austern and Vincent \cite{Aus81}, who  used the post-form 
DWBA. The latter was refined by Kasano and Ichimura \cite{Kas82},  who  
found a formal separation between the EBU and NEB contributions. These 
results were carefully reviewed by Ichimura, Austern and Vincent 
\cite{Ich85} and the model was subsequently referred to as the IAV formalism.  
Later on, Austern {\it et al.} reformulated this theory within a more 
complete three-body model \cite{Aus87}.   

 It is worth noting that the prior-form model of Udagawa and Tamura (UT), on one side, and the post-form DWBA model of Austern and Vincent (AV), on the other side, although formally similar, give different predictions for the NEB part. This led to a long-standing dispute between these two groups, which was finally settled in the referred IAV work \cite{Ich85}, where it was demonstrated that a proper derivation of the prior-form formula gives rises to additional terms not considered by UT. 

Although the comparison of these theories with experimental data showed 
very encouraging results, they have apparently fallen into disuse. Moreover, some of these theories, such as the three-body 
model of  Austern, has never been tested to our knowledge, probably due 
to the computational limitations at that time.   This is in contrast to 
the experimental situation, in which inclusive breakup measurements are 
used for many applications, with both stable and unstable beams. 
Therefore,  it seems timely to reexamine these theories and study their 
applicability to problems of current interest.

The revival and increasing interest on this problem is  evidenced by two recent theoretical works 
on this subject \cite{Car15,Pot15}. Both of them use the IAV model, in DWBA. In Ref.~\cite{Car15}, the authors use the zero-range post-form of this model, whereas in Ref.~\cite{Pot15} the finite-range prior-form version of the model is used instead. Both of them apply the method to deuteron induced reactions, with encouraging results.

In this paper, we revisit also the IAV model,  with special emphasis on the 
calculation of the NEB part, for which we provide a new derivation.  We 
have implemented the DWBA version of this model  both in zero-range and 
in exact finite-range.  To assess the validity of this theory, we have 
performed calculations for several reactions induced by deuterons and, for the first time, the method is applied to 
$^{6}$Li scattering. In both cases, we compare with available data.



The paper is organized as follows. In Sec.~\ref{sec:formalism} 
we give a short overview of the theory, including a new derivation of 
the NEB formula within the IAV model. In Sec.~\ref{sec:calc}, the 
formalism is applied to several inclusive reactions induced by deuterons 
and $^{6}$Li. Finally, in Sec.~\ref{sec:sum} we summarize the main results 
of this work and outline some future developments.

\section{\label{sec:formalism} The Ichimura, Austern, Vincent (IAV) model}
In this section we briefly review the  model of Ichimura, 
Austern and Vincent \cite{Ich85,Aus87}. The final formula obtained in this 
model has been derived in different ways. Here, we closely follow the early 
derivation done by Austern and Vincent \cite{Aus81} because it provides 
an interesting physical insight. 

We write the process under study as
\begin{equation}
a (=b+x) + A \rightarrow b + B^* .
\end{equation}
%
 
\begin{figure}[tb]
\begin{center}
 {\centering \resizebox*{0.95\columnwidth}{!}{\includegraphics{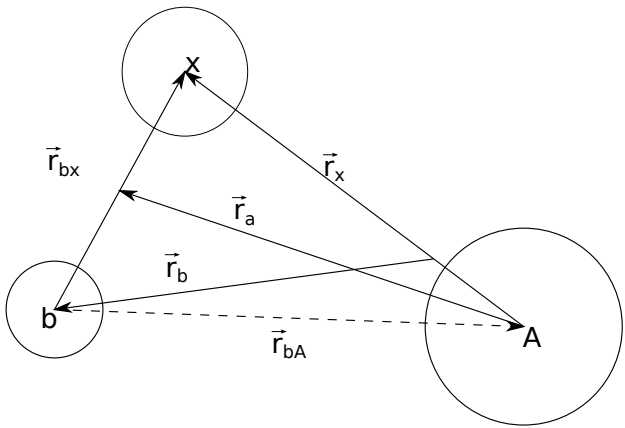}} \par}
\caption{\label{zrcoor}Coordinates used in the  breakup reaction.}
\end{center}
\end{figure}

This process will be described with the  Hamiltonian
\begin{equation}
\label{eq:H3b}
H= K + V_{bx} + U_{bA}(\vecr_{bA}) + H_A(\xi) + V_{xA}(\xi,\vecr_{x}) ,
\end{equation}
where $K$ is the total kinetic energy operator, $V_{bx}$ is the 
interaction binding the two clusters $b$ and $x$ in the projectile $a$,
 $H_{A}(\xi)$ is the Hamiltonian of the target 
nucleus (with $\xi$ denoting its internal coordinates) and $V_{xA}$ and 
$U_{bA}$ are the fragment--target interactions. The relevant coordinates 
are depicted in Fig.~\ref{zrcoor}. Note that the coordinate 
$\vecr_{b}$  connects the particle $b$ with the center of mass (c.m.) of 
the $x+A$ system. 

In writing the Hamiltonian of the system in the form (\ref{eq:H3b}) we 
make a clear distinction between the two cluster constituents; the 
interaction of the fragment $b$, the one that is assumed to be detected 
in the experiment, is described with an optical potential. Non-elastic 
processes arising from this interaction (e.g.~target excitation), 
are included only effectively through $U_{bA}$. 
The particle $b$ is said to act as {\it spectator}.  
 On the other hand, the interaction of 
the particle $x$ with the target retains the dependence of the target degrees of freedom 
($\xi$).

Within the assumed three-body model, and using the post-form representation, the total wave function of the 
system can be written in integral form as
\begin{align}
\Psi(\xi,\vecr_x,\vecr_{b}) & = \left[E^+ - K_b - U_{bB}-H_{B} \right]^{-1} \nonumber \\
            & \times V_\mathrm{post} \Psi(\xi,\vecr_x,\vecr_{b}) ,
\label{eq:Psi3b}
\end{align}
where $E^+=E+i\epsilon$, $\epsilon \rightarrow 0$, $U_{bB}$ is an auxiliary (and, in principle, arbitrary) potential between $b$ and the composite $B$,   $V_\mathrm{post}\equiv V_{bx}+U_{bA}-U_{bB}$ and $H_{B}$ is the Hamiltonian of the $x$+$A$ pair, given by
\begin{equation}
H_B(\xi,\vecr_x)= H_{A}(\xi) + K_x + V_{xA}(\xi,\vecr_{x}) .
\end{equation}
The eigenstates of the target Hamiltonian will be denoted as $\phi^{c}_A(\xi)$, i.e., 
$[H_A(\xi_A) - E_A^c]\phi^{c}_A(\xi)=0$, with $c=0$ corresponding to the target ground state, for which we assume $E_A^0=0$.

We consider now a specific final state of the detected particle $b$, 
characterized by a given final momentum of this fragment ($\vec{k}_b$). The motion of $b$ will be described by a distorted wave
with momentum ${\vec k}_b$, obtained as a solution of the single-channel equation
\begin{equation}
\left[  K_b+ U_{bB}^\dagger-E_b\right]\chi_b^{(-)}(\vec{k}_b,\vecr_{b})=0 .
\end{equation}
 The wave function describing the motion of $x$ after the breakup, that  will be denoted as $Z_x(\xi,\vecr_x)$, can be 
obtained projecting the total wave function [Eq.~(\ref{eq:Psi3b})] onto this particular state of the $b$ particle, i.e.,%
\footnote{Note that this function will also depend on $\vec{k}_a$, which indicates the direction of the incident beam. Because this direction is fixed, this dependence will be omitted for simplicity of the notation.}
\begin{align}
 \zbx & \equiv ( \chi_b^{(-)} | \Psi \rangle  \nonumber \\
  & =  \left[E^+ - E_b-H_{B} \right]^{-1} (\chi_b^{(-)}| V_\mathrm{post}|\Psi\rangle ,
\label{eq:zx_int}
\end{align}
where the round bracket denotes integration over 
$\vecr_{b}$ only.
The last equation can be also written in differential form as
\begin{equation}
\label{eq:zeq}
\left[E^+ -E_b-H_{B} \right] \zbx=  (\chi_b^{(-)} | V_\mathrm{post}|\Psi\rangle .
\end{equation}
The source term of this equation involves the exact and hence unknown 
wave function $\Psi$ which, in actual calculations, must be approximated 
by some calculable form. For example, in DWBA, 
one  assumes the factorized form
\begin{equation}
\label{eq:psi_dwba}
\Psi(\xi,\vecr_x,\vecr_{b}) \approx \phi^0_A(\xi) \phi_a(\vec{r}_{bx}) 
\chi^{(+)}_{a}(\vec{k}_a,\vec{r}_a) ,
\end{equation}
where $\phi_a(\vec{r}_{bx})$ is the projectile 
ground-state wave function and $\chi^{(+)}_{a}(\vec{k}_a,\vec{r}_a)$ is a distorted 
wave describing the $a+A$ motion in the incident channel. In practice, 
the latter is commonly replaced by the solution of some optical potential describing  $a+A$ elastic scattering. 
Austern {\it et al.} \cite{Aus87} proposed also the three-body approximation 
\begin{equation}
\label{eq:psi_3b}
\Psi(\xi,\vecr_x,\vecr_{b}) \approx \phi^0_A(\xi) \Psi^{\mathrm{3b}}(\vecr_x,\vecr_{b}) ,
\end{equation}
where $\Psi^{\mathrm{3b}}$ is a three-body wave function for the three 
fragments ($x$+$b$+$A$) and contains, in addition to the $b+x$ ground 
state, contributions from $b+x$ inelastic scattering and breakup.

 It is worth noting that, either in the approximation (\ref{eq:psi_dwba}) or in (\ref{eq:psi_3b}), the three-body wave function does not contain explicitly excited states of $A$. Thus, in the IAV model, the NEB  can be viewed as a two-step process in which the first step is the dissociation of the projectile, leaving the target in the ground state, while the second step is the absorption of $x$ or the excitation of $A$.

A possible procedure to solve Eq.~(\ref{eq:zeq}) is to expand the 
function $Z_x$ in a complete set of $x+A$ states, i.e., 
\begin{equation}
\label{eq:zexpand}
\zbx = \sum_{c} \psi_x^{c}(\vec{k}_b,\vecr_x) \phi^{c}_A(\xi) ,  
\end{equation}
where $\psi_x^{c}(\vec{k}_b,\vecr_x)$ describes the $x-A$ relative motion when the 
target is in the state $c$. 
The  expansion (\ref{eq:zexpand}) can be inserted into Eq.~(\ref{eq:zeq}), giving rise to 
a set of coupled equations for the unknown functions $\psi_x^{c}(\vec{k}_b,\vecr_x)$.

This approach will be in general unpractical because the expansion (\ref{eq:zexpand})  involves 
a very large number of final states. If one is not interested in the description 
of the transition to specific $x+A$ states, but rather in their summed contribution, 
one can proceed as follows. Following Feshbach, the $\zbx$ function is decomposed as
\begin{equation}
\zbx = {\cal P} Z_x + {\cal Q} Z_x ,
\end{equation} 
where ${\cal P}$ is the projector operator onto the target ground state and 
${\cal Q}= 1 - {\cal P}$. From Eq.~(\ref{eq:zexpand}) we see 
that ${\cal P} Z_x = \psix \phi^{0}_A(\xi)$.   
The function $\psix$, which describes the $x+A$ relative 
motion when the target is in the ground state, verifies the equation
\begin{equation}
\label{eq:pz}
(E^+_x - K_x - {\cal U}_x)  \psix =  \rho(\vec{k}_b,\vec{r}_x) 
\end{equation}
with $E_x=E-E_b$, $\rho(\vec{k}_b,\vec{r}_x) \equiv  (\chi_b^{(-)}| V_{bx}|\Psi \rangle$ is the so-called source term,
and  ${\cal U}_x$ the formal optical model potential describing $x+A$ elastic scattering. Explicitly,
\begin{equation}
{\cal U}_x =\langle \phi^0_A |V_{xA} + V_{xA} Q [E^+ - E_b - H_{QQ}]^{-1}  V_{xA}| \phi^0_A \rangle  ,
\end{equation}
where $H_{QQ} \equiv {\cal Q} H_{B} {\cal Q}$.  The formal potential 
${\cal U}_x$ is a complicated non-local, angular- and energy-dependent 
object.  However, as done in two-body scattering problems, it can be 
approximated by some energy-averaged (possibly local) potential or by 
some phenomenological representation  (denoted $U_x$ hereafter) with 
parameters adjusted to  describe $x+A$ elastic scattering. 

Note that Eq.~(\ref{eq:pz}) is formally analogous to the inhomogeneous 
equation appearing in DWBA and CCBA calculations between bound states, as 
formulated in the {\it source term  method} of Ascuitto and Glendenning 
\cite{Asc69}, and used by several coupled-channels codes \cite{Thom88}.  

\subsection{Separation of elastic and nonelastic breakup}
In their original paper, Austern and Vincent provide only the total 
inclusive cross section. Later on, Kasano and Ichimura \cite{Kas82} 
showed that this expression can be formally decomposed into two pieces, 
corresponding to the elastic breakup (EBU) and non-elastic breakup 
(NEB) contributions.

Here, we present an alternative derivation of these  formulas, which 
exploits the aforementioned analogy of Eq.~(\ref{eq:pz}) with that 
found in the DWBA and CCBA formalisms. This equation is to be solved with 
purely outgoing boundary conditions (since there are no incoming waves 
in the $x-A$ channel), that is,
\begin{equation}
\label{eq:psix_asym}
 \psix  \rightarrow f(\vec{k}_b,\hat{r}_x) \frac{ e^{i k_x r_x }}{r_x} . 
\end{equation}
 The function $f(\vec{k}_b,\hat{r}_x)$ depends, in addition to the direction of $\vec{k}_b$, on the angular part of $\vec{r}_x$. Asymptotically, when $r_x$ is large, the position vector $\vec{r}_x$ becomes parallel to the momentum $\vec{k}_x$ and we may write $f(\vec{k}_b,\hat{r}_x)  \rightarrow f(\vec{k}_b,\vec{k}_x)$. We therefore recognize $f(\vec{k}_b,\vec{k}_x)$ as the scattering amplitude for the elastic breakup process, 
and its square is proportional to the differential cross section for the 
detection of the $x$  particle in the direction of $\vec{k}_x$, and the 
$b$ particle in the direction $\vec{k}_b$. To obtain this amplitude, one 
can proceed in two different ways. One possibility is to integrate the 
differential equation (\ref{eq:pz}) and, at a sufficiently large distance 
(beyond the range of the short-ranged potentials), equate the solution to 
the asymptotic form (\ref{eq:psix_asym}), from which the scattering 
amplitude can be obtained. A second approach to solve Eq.~(\ref{eq:pz}) is to use integral methods 
(Green function) techniques. This gives a 
closed-form expression for the scattering amplitude,  
\begin{equation}
f(\vec{k}_b,\vec{k}_x) = - \frac{\mu_{x}}{2 \pi \hbar^2} \langle \chi^{(-)}_x  \, \chi^{(-)}_b | V_{bx} | \Psi^{3b} \rangle ,
\end{equation}
where $\mu_{x}$ is the reduced mass of the $x+A$ system and the distorted wave $\chi^{(-)}_x(\vec{k}_x,\vec{r}_x)$ is a solution of the homogeneous part of equation Eq.~(\ref{eq:pz}), i.e.,
\begin{equation}
\left[ K_x+U_x^{\dagger}-E_x \right]\chi^{(-)}_x(\vec{k}_x,\vecr_x)=0 ,
\end{equation}
whose solution consists of  a plane wave of momentum $\vec{k}_x$
plus an ingoing spherical wave.

The corresponding differential cross section,  for a final differential volume in momentum space, is given by%
\footnote{Note that the factor $(2 \pi)^{4}$ of Ref.~\cite{Gol04} is 
replaced here by a $(2 \pi)^{-5}$ factor, consistent with our definition of the amplitude for the plane waves as $e^{i \vec{k} \vec{r}}$.} %
(c.f., for instance, Eq.~(5.36) of Ref.~\cite{Gol04})
\begin{equation}
d\sigma=  \frac{(2 \pi)^{-5}}{\hbar v_i}\int d\vec{k}_x d\vec{k}_b d\vec{k}_A ~\delta(E_f - E_i) \delta(\vec{P}_f-\vec{P}_i) |T_{fi}|^2 ,
\end{equation}
where $T_{fi}$ is the usual transition amplitude (or T-matrix), which is 
related to the scattering amplitude by $f=-(\mu_{x} / 2 \pi \hbar^2) T_{fi}$.
In the c.m.\ frame, 
$\vec{P}_i=0$. Also, the target momentum ($\vec{k}_A$) is not measured, so we can 
integrate over it, making use of the momentum-conserving delta function, 
\begin{equation}
\label{eq:dsig_kxkb}
d\sigma= \frac{(2 \pi)^{-5}}{\hbar v_i} \int d\vec{k}_x d\vec{k}_b  \delta(E_f - E_i)|T_{fi}|^2 .
\end{equation}
The element $d\vec{k}_b$ is conveniently expressed in terms of energy 
and solid angle elements using $d\vec{k}_b= (2\pi)^3 \rho_b(E_b) d\Omega_b dE_b$, where $\rho_b(E_b)=k_b \mu_{b} /((2\pi)^3\hbar^2)$
is  the density of states.\footnote{These expressions result from 
$N(k) d\vec{k}_b= \rho_b(E_b) d\Omega_b dE_b$, where $N(k)$ is the number 
of states in the differential volume $d\vec{k}_b$, which is determined 
from $\langle \vec{k}| \vec{k}'\rangle = \delta(\vec{k}-\vec{k}')/N(k)$. 
In our case, $\langle \vec{k} | \vec{k}  \rangle =(2 \pi)^3\delta(\vec{k} - \vec{k}')$, and hence $N(k)=(2\pi)^{-3}$.
} Using this in Eq.~(\ref{eq:dsig_kxkb}),
\begin{equation}
{d\sigma}= \frac{(2 \pi)^{-2}}{ \hbar v_i} \int \delta(E_f - E_i)|T_{fi}|^2  \rho_b(E_b) {d\Omega_b dE_b} d\vec{k}_x .
\end{equation}
The double differential cross section with respect to the energy and the scattering angle of  $b$ is therefore given by
\begin{equation}
\left . \frac{d^2\sigma}{d\Omega_b dE_b}  \right |_\mathrm{EBU} = \frac{(2 \pi)^{-2}}{ \hbar v_i} \rho_b(E_b) \int \delta(E_f - E_i)|T_{fi}|^2  d\vec{k}_x ,
\label{eq:iaveb}
\end{equation}
which coincides with the result of Austern {\it et al.} (Eq.~(8.15) of Ref.~\cite{Aus87})  noting that $\int d\vec{k}_x \rightarrow (2\pi)^3 \sum_{\vec{k}_x}$.

Although it is not the purpose of the present work, we note also that the 
previous expression can be used to compute the fully exclusive cross section, 
with respect to the  angles and energies of $b$ and $x$. For that, we use again  
$d\vec{k}_x= (2\pi)^3 \rho_x(E_x) d\Omega_x dE_x$ and use the energy-conserving delta function, resulting
\begin{equation}
\left . \frac{d^2\sigma}{d\Omega_b dE_b d\Omega_x}  \right |_\mathrm{EBU} = \frac{2 \pi}{ \hbar v_i} 
  \rho_b(E_b) \rho_x(E_x)  |T_{fi}|^2  .
\end{equation}
%


To obtain the expression for the NEB component we make use 
of the {\it coupled-channels optical theorem} recently formulated by 
Cotanch \cite{Cot10}. This work generalizes the well-known optical theorem 
to the multichannel case. 
If $\chi_i$ is the channel wave function and 
$W_i$ the diagonal imaginary part for this channel, the contribution to 
the absorption in this particular channel is given by \cite{Cot10}
\begin{equation}
\sigma^i_\mathrm{abs}=-\frac{2}{\hbar v_{el}} \langle \chi_i | W_i | \chi_i \rangle ,
\end{equation}
where $v_{el}$ is the projectile--target relative velocity in the 
incident (elastic) channel. 

We may use this result to calculate the NEB contribution by noting that 
the latter is nothing but the absorption occurring in the $x+A$ 
channel. The channel wave function is given by $\psix$, which is a 
solution of Eq.~(\ref{eq:pz}). 
Since Eq.~(\ref{eq:pz}) corresponds to a definite energy and direction of the $b$ particle, we consider the 
differential cross section corresponding to a range of the outgoing  
momenta of $b$,
\begin{equation}
d^2\sigma = -\frac{2}{\hbar v_{i}} \langle \psi^{0}_x | W_x | \psi^{0}_x \rangle  \, N(k_b) \, d\vec{k}_b ,
\end{equation}
with $W_x \equiv \Im[U_x]$. Transforming the element of momentum into 
energy and solid angle elements, we get the double differential cross section
\begin{equation}
\label{eq:iav}
\left . \frac{d^2\sigma}{dE_b d\Omega_b} \right |_\mathrm{NEB} = -\frac{2}{\hbar v_{i}} \rho_b(E_b)  \langle \psi^{0}_x | W_x | \psi^{0}_x \rangle   .
\end{equation}

This result was obtained, by different arguments, by Kasano and Ichimura 
\cite{Kas82}. A similar result was also obtained by Hussein and McVoy 
\cite{HM85}. The alternative derivation presented here, 
based upon the generalized optical theorem, 
provides a clear interpretation of this term, as the flux leaving the 
$x+A$ channel following the breakup of the projectile into $b$+$x$. 

To recapitulate, in the IAV model, the breakup can be viewed as a 
two-step process. The first step corresponds to the dissociation of the 
projectile ($a$) into the fragments $b$ and $x$, leaving the target in the 
ground state. The subsequent motion of the participant particle ($x$) is 
described by the function $\psix$, which is the solution of 
the inhomogeneous Eq.~(\ref{eq:pz}).  This particle can then be scattered 
elastically by the target or can interact non-elastically (for example, 
excite the target or fuse with it).  The former corresponds to the EBU 
part of the inclusive breakup cross section whereas these non-elastic 
processes,  corresponding to the second step in this two-step picture,  
yield the NEB contribution. Quantitatively, this contribution is obtained as the expectation value of  
 $\Im[U_x]$ in the state  $\psix$ [Eq.(\ref{eq:iav})]. 
 Note that, since this function  depends on the final state of the 
 {\it spectator} particle ($b$), the NEB expression (\ref{eq:iav}) yields the absorption for each final state of $b$.


\subsection{Practical implementation of the IAV model}
The IAV formula for NEB breakup, Eq.~(\ref{eq:iav}), has a deceptively 
simple form. The function $\psi^{0}_x$ must be first calculated from the 
inhomogeneous Eq.~(\ref{eq:pz}), whose source term contains the 
three-body wave function $\Psi^{3b}$, which is a complicated object by 
itself. Furthermore, this equation must be solved for each outgoing 
energy and angle of $b$ covering the range of interest.

For these reasons, practical implementations of this theory have 
resorted to additional approximations. Standardly, all these 
applications rely on the DWBA approximation of the incident channel [that is, Eq.~(\ref{eq:psi_dwba})], 
rather than on a three-body model [Eq.~(\ref{eq:psi_3b})]. Even at the DWBA level, 
Eq.~(\ref{eq:pz}) is not trivial. Usually, a partial wave decomposition 
of the scattering waves appearing in Eq.~(\ref{eq:pz}) will be used and 
this means that a large number of angular momenta for the $a+A$, $x+A$, 
and $b+B$ distorted waves will be required for convergence of the 
calculated cross sections. In addition, the right-hand-side of this 
equation contains  non-local kernels (similar  to those appearing 
in DWBA calculations between bound states, but involving a larger number of 
angular momenta).  Consequently, in addition to the DWBA approximation, 
most of the existing calculations of this kind have been done in the 
zero-range  (ZR) approximation.  

To assess the validity of this approximation, we have performed calculations 
with both the ZR and exact finite-range (FR) calculations. The detailed 
formulas for the NEB cross sections in these two approximations are given in the Appendices.

Another difficulty arising in solving Eq.~(\ref{eq:pz}) are the well-known convergence problems of the post-form DWBA formula when applied to breakup reactions. 
This is because $\chi^{(-)}_b$, being  a scattering state, 
will be infinitely oscillatory and the operator in the matrix element 
$V_{bx}$ and the initial state ($\psi_a$ in DWBA) depend on the 
$\vec{r}_{bx}$ coordinate and hence there is no natural cutoff in the 
$\vec{r}_{b}$ integration. As a consequence, the source term has infinite range.
To overcome this problem, Huby and Mines 
\cite{Hub65} and Vincent \cite{Vin68} multiply the source term by an 
exponential convergence factor, that damps the contribution of the 
integral at large distances. Alternatively, following Vincent and Fortune 
\cite{Vin70}, one may use the integration in the complex plane.  
Here, we adopt a different procedure. Following Thompson \cite{Tho11}, 
we consider energy {\it bins} for the $b$ distorted waves.  For this, 
the scattering states are first expanded in partial waves (see Appendix~\ref{sec:zr}), and the radial 
coefficients, $R_{\ell_b}(r_{b},k_b)$ are then averaged over  small 
energy or momentum intervals, i.e.,
\begin{equation}
\label{eq:Rbin}
\bar{R}_{\ell_b}(r_{b},k^i_b) =N \int_{k^i_b-\Delta {k_b}/2}^{k^{i}_b+\Delta {k_b}/2} dk_b ~ R_{\ell_b} (r_{b}, k_b)   ,
\end{equation}
where  $\Delta k_b$ is the bin width, $k^i_b$ the central momentum of the bin
 and $N$ is a normalization constant. 
The resulting bin wave function is square-integrable and thus leads to 
convergent results when it is used in the source term of Eq.~(\ref{eq:pz}). 

The formulas discussed in this section are applied to specific cases in the following section.

\section{\label{sec:calc} Calculations}
In this  section, we present calculations for several reactions  induced 
by deuterons and $^6$Li projectiles, and compare the calculated inclusive 
cross sections with experimental data, in order to assess the validity of 
the theory. In all cases, we compute the separate contributions for the 
elastic (EBU) and non-elastic (NEB) breakup cross sections. For the former,  
we use the CDCC formalism, using the coupled-channels code {\sc fresco} \cite{Thom88}. 
This  permits to treat the EBU to all orders, and should be equivalent 
to the post-form three-body model of Austern {\it et al}. For the NEB part, 
we use the DWBA version of Eq.~(\ref{eq:iav}).  
We have also tested the accuracy of the ZR approximation 
in the NEB formula, by comparing ZR with FR calculations. 

\subsection{Application to $(d,pX)$}
There is a large body of exclusive and inclusive breakup data for 
deuteron-induced reactions. We have considered the inclusive $(d,pX)$ data 
for the reactions $d$+$^{93}$Nb at $E_d=25.2$~MeV from Ref.~\cite{Pam78}, 
and $d$+$^{58}${Ni} at 80 and 100 MeV from Refs.~\cite{Wu79,Ridikas00}.

\begin{figure}[tb]
\begin{center}
 {\centering \resizebox*{0.95\columnwidth}{!}{\includegraphics{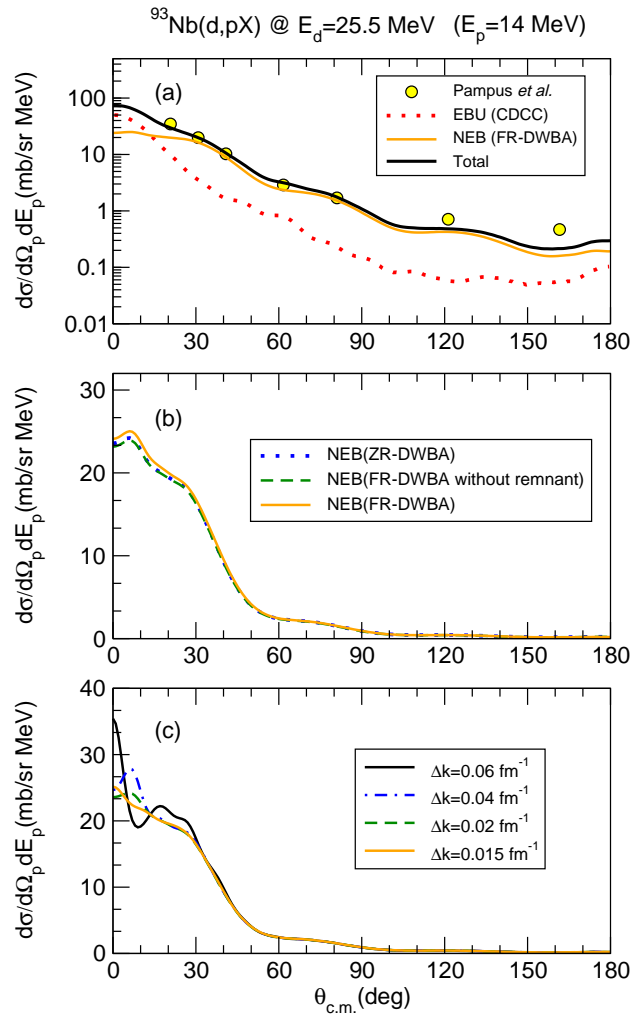}} \par}
\caption{\label{nb93dp_dsdw}(Color online) a) Experimental and calculated double differential cross section, as a 
function of the proton scattering angle, for the protons emitted in the 
$^{93}$Nb($d$,$p$X) reaction with an energy of 14 MeV, and a deuteron incident energy of $E_d=25.5$ MeV. The dotted, thin solid and thick solid lines are the elastic breakup  (CDCC),  the non-elastic breakup (FR-DWBA)  and their incoherent sum, respectively. Experimental data are from Ref.~\cite{Pam78}; b) Non-elastic breakup angular distribution calculated with ZR-DWBA (dotted), FR-DWBA without remnant (dashed) and full FR-DWBA (solid line); c) Convergence of the NEB calculation with respect to the bin width, $\Delta k_b$, used for  the $b$ distorted waves. See text for details.}
\end{center}
\end{figure}

The data for $d$+$^{93}$Nb were already analyzed in Ref.~\cite{Pam78}, 
using the so-called surface approximation, in Ref.~\cite{Kas82}, 
using the zero-range version of the post-form DWBA formula discussed here, and in Ref.~\cite{Pot15}, using the prior form of the DWBA IAV model.  
These calculations give a reasonable account of the experimental data.

 In the  CDCC calculations \cite{Aus87} the deuteron breakup is treated 
 as inelastic excitations to the $p$-$n$ continuum. This  continuum is 
 truncated at a maximum excitation energy, and discretized in energy bins. 
 For the present case, the $p$-$n$ states were included for $\ell=0-4$ 
 partial waves, and up to a maximum excitation energy of 20~MeV. For the 
 $p$-$n$ interaction, we considered the simple Gaussian form of Ref.~\cite{Aus87}. 
 The proton-target and neutron-target interactions were adopted from the 
 global parametrization of Koning and Delaroche (KD) \cite{KD02}, omitting 
 the spin-orbit term, and evaluated at half of the deuteron incident energy. 
 In the CDCC method, the breakup cross sections are calculated in terms 
 of the c.m.\ scattering angle and excitation energy of the $p$-$n$ system. 
 Therefore, to compare with the proton inclusive data, these breakup 
 cross sections must be converted to the proton energy and scattering 
 angle, making use of the appropriate kinematical transformation. 
 This was done with the formalism and codes developed  in Ref.~\cite{Tostevin01}. 
 For the NEB calculations, we use also the KD parametrization for the 
 proton-target and neutron-target interactions, but evaluated at the 
 corresponding proton ($E_p$) and neutron ($E_n$) energies. In DWBA,   
 one needs also the incoming channel optical potential ($d$+$^{93}$Nb), 
 which was taken from Ref.~\cite{An06}.  For the ZR-DWBA calculations 
 we used the zero-range constant   $D_0=125~\mathrm{MeV}\cdot\mathrm{fm}^{3/2}$, and included the finite-range correction factor 
 (see, e.g., Refs.~\cite{But64,satchler83} and Appendix \ref{sec:zr}).

In Fig.~\ref{nb93dp_dsdw}(a)  we compare the experimental  \cite{Pam78} and calculated  inclusive double differential 
cross section,  $\mathrm{d}^2\sigma/\mathrm{d}
E_p\mathrm{d}\Omega_p$, corresponding to a proton energy of $E_p=14$ MeV.   
The dotted line is the EBU calculation (CDCC), which is found to 
underestimate the data at all angles.  The thin solid line is the FR-DWBA calculation for the NEB part 
(see Appendix~\ref{sec:fr}).  The thick solid line is the sum of the EBU  and NEB 
contributions. Except at very large angles, 
it is found to explain satisfactorily the data. It is seen that, except 
for the smallest angles, the inclusive breakup cross section is largely 
dominated by the NEB contribution. Our results are consistent with those 
reported in Refs.~\cite{Pam78} and \cite{Kas82}.   In Fig.~\ref{nb93dp_dsdw}(b), we compare several approximations for the numerical evaluation of the NEB cross section. The dotted line is the ZR-DWBA calculation, including nevertheless the finite-range correction $\Lambda(r_{x})$ (see Appendix  ~\ref{sec:zr}). The dashed line is the FR-DWBA calculation,  omitting the remnant term in the transition operator (i.e., using $V_\mathrm{post} \approx V_{pn}$). Finally, the solid line is the full FR-DWBA calculation.  We find that the ZR calculation (with finite-range correction) provides a very accurate result in the present reaction,  thus supporting the  validity of this approximation in this case. Further, we see that the non-remnant term has a very small effect, and can be also safely ignored in the FR calculation.  

In order to obtain meaningful results, the calculated observables must converge as the bin width $\Delta k_b$  is progressively decreased [c.f.~Eq.~(\ref{eq:Rbin})]. This is verified in Fig.~\ref{nb93dp_dsdw}(c) for the present case, where we show the calculated NEB angular distribution for different values of $\Delta k_b$. Although the rate of convergence was found to be small, it is seen that for $\Delta k_b\approx 0.02$~fm$^{-1}$ the calculations are well converged for the full angular range. A similar convergence study was done in the other calculations presented below.

\begin{figure}[tb]
\begin{center}
 {\centering \resizebox*{0.95\columnwidth}{!}{\includegraphics{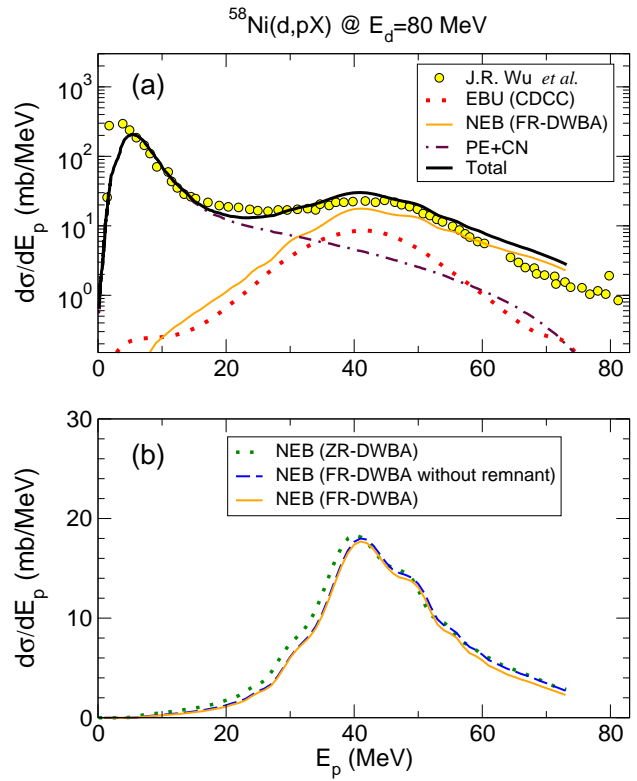}} \par}
\caption{\label{ni58dp_dsde}(Color online)(a) Experimental and calculated angle-integrated proton differential cross section, as a 
function of the outgoing proton energy in the LAB frame, for the 
$^{58}$Ni($d$,$pX$) reaction at $E_d=80$ MeV.  The dotted and thin solid lines are the EBU and NEB contributions, calculated with CDCC and FR-DWBA, respectively. The dot-dashed line is the contribution coming from pre-equilibrium and compound nucleus  \cite{Wang11}. The thick solid line is the incoherent sum of the three contributions.  Experimental data are from 
Ref.~\cite{Wu79}. (b) Non-elastic breakup  calculated with ZR-DWBA (dotted), non-remnant FR-DWBA (dashed), and full FR-DWBA (solid) formulas.}
\end{center}
\end{figure}

We present now the results for the $^{58}$Ni($d$,$pX$) reaction at 80 and 
100 MeV, and compare with the data from Refs.~\cite{Wu79,Ridikas00}. 
These data have been also analyzed in Refs.~\cite{Ye11,Wang11,Nakayama14}, 
using the CDCC method for the EBU part, and the semi-classical Glauber 
approach for the NEB part. In our CDCC calculations, the proton-target 
and neutron-target interactions are obtained  again from the 
Koning-Delaroche parametrization, and we employed the same $p$-$n$ interaction  
used in the $d$+$^{93}$Nb calculations. For the  $p$-$n$ continuum we 
considered the partial waves  $\ell=0-6$, and  excitation energies up to 
50~MeV and 90~MeV for the data at $E_d=80$~MeV ad $E_d=100$~MeV, respectively.  
For the NEB calculations, the $d$+$^{58}$Ni potential was taken from Ref.~\cite{An06}.

In Fig.~\ref{ni58dp_dsde}, we present the angle-integrated energy 
differential cross section   at $E_d$=80~MeV ($d\sigma/dE_p$).  In Fig.~\ref{ni58dp_dsde}(a), the dotted and thin solid 
lines correspond to the EBU (CDCC) and NEB (FR-DWBA) calculations.
It is seen that the NEB contribution is much larger 
than the EBU part. Both  distributions show a bell-shaped 
behavior, with a maximum around half of the deuteron energy. However, 
it is observed that the sum of these two contributions cannot explain the  
experimental yield at small proton energies. As shown in Ref.~\cite{Wang11}, 
these low-energy protons come mainly from compound nucleus followed by 
evaporation and pre-equilibrium. Since these processes are not accounted 
for by the present formalism, in this work we have adopted the estimate  
done in Ref.~\cite{Wang11} (dot-dashed line in Fig.~\ref{ni58dp_dsde}(a)). 
The total inclusive cross section, including this contribution 
(thick solid line) reproduces reasonably well the shape and magnitude 
of the data. Note that,  protons with energies larger than $\sim$74~MeV, 
correspond to bound states of the neutron-target system and they are 
associated with a stripping mechanism. This contribution could be accommodated 
in the present formalism solving Eq.~(\ref{eq:pz}) for $E_x<0$ and with boundary conditions 
appropriate for bound states instead of outgoing boundary conditions. 
 Further, for high-lying bound excited states, were the density of levels will be very 
high, one may use the ideas of Udagawa and co-workers of extending the 
complex potential to negative energies to describe the spreading of 
single-particle states \cite{Uda87,Uda89}. These extensions go however beyond the scope of the present work.   

In  Fig.~\ref{ni58dp_dsde}(b), we compare 
different approximations for the transition amplitude used in the NEB calculation, namely, ZR-DWBA (dotted), 
FR-DWBA with no remnant (dashed) and full FR-DWBA (solid). As in the previous case, the ZR-DWBA and FR-DWBA calculations 
agree very well for proton energies around and above the maximum, although 
some small differences are visible. The effect of the remnant term is again found to be very small. 

\begin{figure}[tb]
\begin{center}
 {\centering \resizebox*{0.95\columnwidth}{!}{\includegraphics{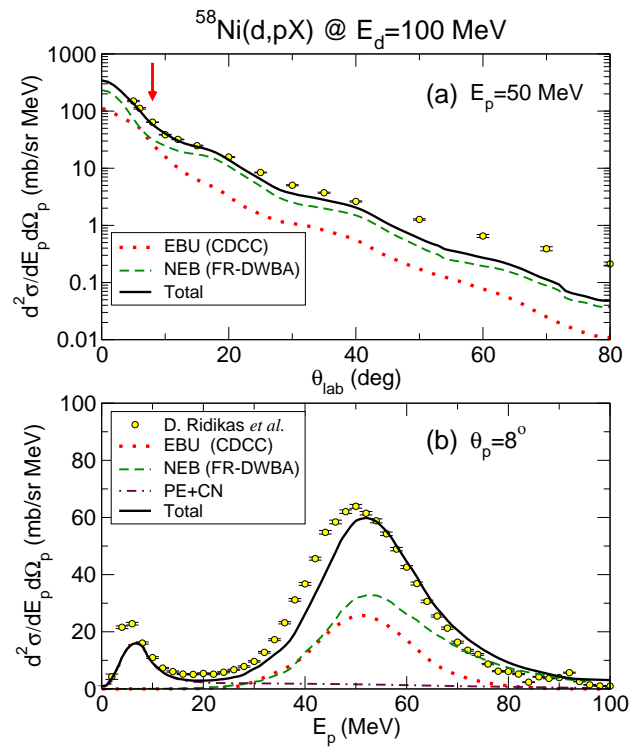}} \par}
\caption{\label{ni58dp_dsdwde}(Color online) Double differential cross section of 
protons emitted in the $^{58}$Ni($d$,$pX$) reaction at $E_d=100$ MeV in the laboratory
frame. (a) Proton angular distribution for a fixed proton energy of $E_p=50$~MeV. 
(b) Energy distribution for protons emitted at a laboratory angle of 8$^\circ$ (arrow in top figure). 
The meaning of the lines is the same as in Fig.~\ref{ni58dp_dsde}, and are also indicated by the labels. 
Experimental data are from Ref.~\cite{Ridikas00}.}
\end{center}
\end{figure}

We finally present the results for the $d$+$^{58}$Ni reaction at 100~MeV. 
This is shown in Fig.~\ref{ni58dp_dsdwde}, where 
the top panel contains the experimental and calculated proton angular distributions 
for protons detected at 50~MeV in the laboratory frame, 
and the bottom panel shows the  energy distribution for the protons scattered 
at $8^\circ$ in the laboratory frame. Again, it is seen that the inclusive 
breakup is dominated by the NEB contribution in the full angular range,  
particularly at large scattering angles. As in the 80~MeV case, both the EBU and 
NEB contributions exhibit bell-shaped distributions, with a maximum around 
$\approx E_d/2$. On the other hand, the protons coming from compound nucleus 
and pre-equilibrium dominate the low-energy region. Except for some 
underestimation of the cross section at the maximum, the agreement 
between the theory and the data is rather satisfactory. 

\subsection{Application to  ($^6$Li,$\alpha X$) }
\begin{figure}[tb]
\begin{center}
 {\centering \resizebox*{0.95\columnwidth}{!}{\includegraphics{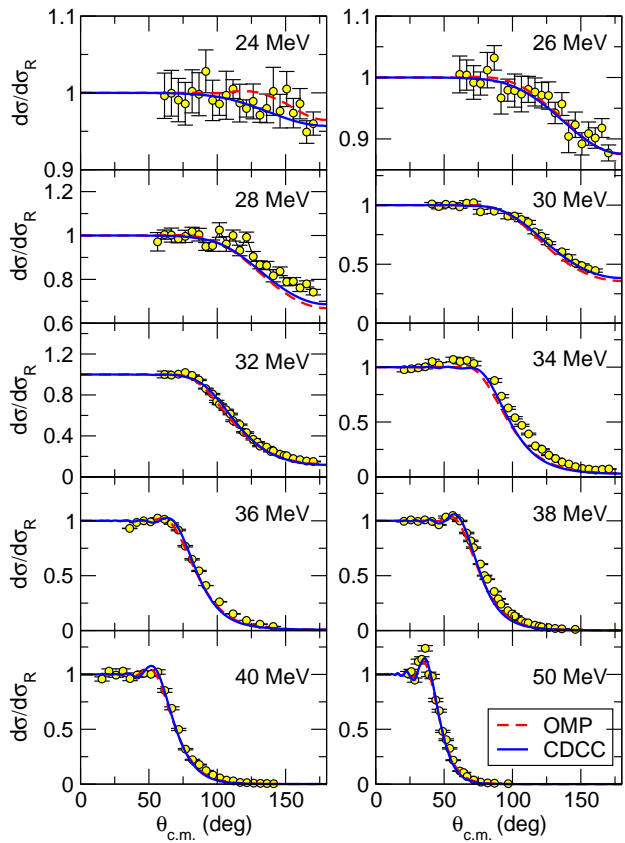}} \par}
\caption{\label{6liela}(Color online) Elastic scattering of  $^6$Li+$^{209}$Bi at different incident 
energies. The solid and dashed  lines are, 
respectively, the CDCC calculation and the optical model calculation with the optical potential from \cite{Cook82}. The experimental data are from Ref.~\cite{Santra11} .}
\end{center}
\end{figure}

\begin{figure}[tb]
\begin{center}
 {\centering \resizebox*{0.95\columnwidth}{!}{\includegraphics{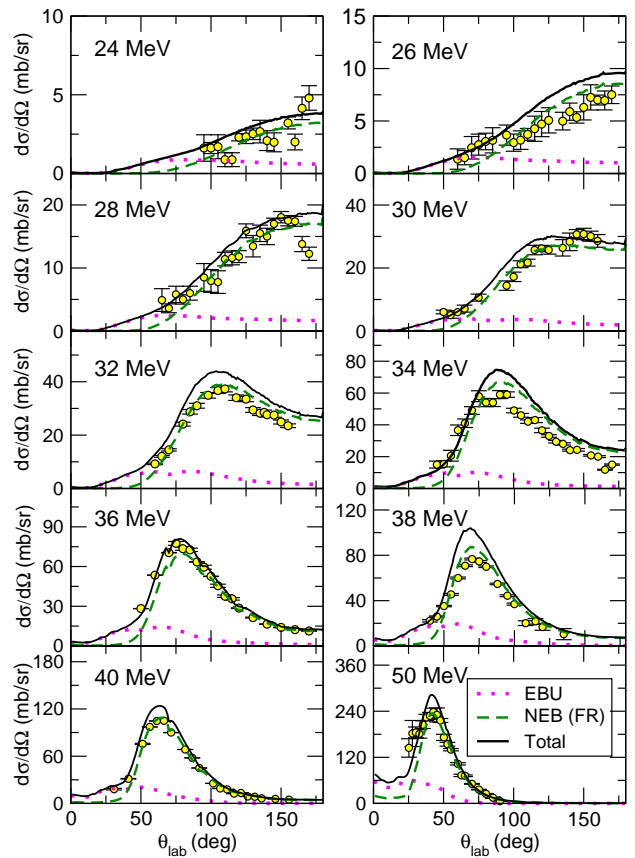}} \par}
\caption{\label{li6bi_dsdw}(Color online) Angular distribution of $\alpha$ particles produced in the reaction $^6$Li+$^{209}$Bi at the incident energies indicated by the labels. The  dotted, dashed and solid lines correspond to the EBU (CDCC), NEBU (FR-DWBA) and their sum, respectively. Experimental data 
are from Ref.~\cite{Santra12}.}
\end{center}
\end{figure}
As a second example, we consider the $\alpha$ production following the 
breakup of the weakly-bound nucleus $^{6}$Li. The understanding of the 
large $\alpha$ yields observed in reactions with $^6$Li has been subject 
of many studies \cite{Pfeiffer73,Cas80,Kelly00,Pak03,Sig03,Souza09,Kumawat10,Santra12,Pra13}. 
These works have shown (see e.g.~Refs.~\cite{Kumawat10,Sig03}) that the 
total exclusive cross sections ($\alpha$+$d$ and $\alpha$+$p$) are much 
smaller than the total $\alpha$ production cross section. Consequently, 
the $\alpha$ inclusive cross sections are largely underestimated by CDCC calculations. 
Furthermore, some of these works have shown that the {\it total} fusion cross section 
of these reactions is significantly enhanced due to partial fusion of the projectile, usually referred to as incomplete fusion (ICF) \cite{Canto06}.
The calculation of ICF cross sections from a purely quantum mechanical framework is still a challenging problem \cite{Canto06,Mar14}. Since the ICF is part of the NEB cross section, the inclusive breakup model considered in this work, might provide useful starting point to tackle this problem.   However, one has to bear in mind that the NEB cross section will contain, in addition to ICF contributions, other contributions, such as breakup accompanied by target excitation, without absorption of any of the fragments. These contributions should be subtracted from the total NEB cross section in order to extract the ICF part. Work in this direction is in progress and the results will be presented elsewhere. Here, we focus on the calculation of the total inclusive cross sections.   

For that purpose, we have considered the $^{6}$Li+$^{209}$Bi reaction at 
several bombarding energies between 10 and 50 MeV, for which experimental 
data exist \cite{Santra12}. The nominal Coulomb barrier for this system 
is around 30.1~MeV \cite{Das02}, so these data cover energies below and 
above the barrier. The $^{6}$Li nucleus is treated in a two-cluster model ($\alpha$+$d$). 
CDCC calculations based on this model have been performed for many $^{6}$Li 
induced reactions. In order to reproduce the elastic data, these calculations usually
require a reduction of the imaginary part of the fragment-target 
interactions \cite{Hir91,Bec07,San09}. On the other hand, four-body CDCC 
calculations, based  on a more realistic three-body model of 
$^{6}$Li ($\alpha$+$p$+$n$), are able to describe the elastic data for 
$^{6}$Li+$^{209}$Bi without any readjustment of these potentials \cite{Wat12}, 
thus suggesting that the need for a reduced absorption   is related to the limitations of this two-body model 
for $^{6}$Li.  Since the inclusive formulas considered in this work are 
based on a two-body model of the projectile, we perform our calculations 
with the $\alpha$+$d$ model, and allow for the same kind of 
renormalization prescribed in previous works.   

For that, we first study the elastic scattering within the CDCC framework. 
These calculations include $s$-wave  ($J^\pi=1^+$),  $p$-wave 
($J^\pi=0^-,\ 1^-,\ 2^-$), and $d$-wave  ($J^\pi=1^+,\ 2^+,\ 3^+$) continuum states. 
For the $d$ wave, we make a finer division of bins in order to describe the 
$^6$Li resonant states at 2.186 MeV ($J^\pi=3^+$),  4.31 MeV ($J^\pi=2^+$) 
and  5.7 MeV ($J^\pi=1^+$). 
For the $\alpha+d$ ground state we used a Woods-Saxon well with 
$V_0=78.46$ MeV, $r_0=1.15$ fm, and $a=0.7$ fm \cite{Nishioka84}. We 
used a second Woods-Saxon well to describe the $p-$ and $d-$wave states 
with parameters $V_0=80.0$ MeV, $r_0=1.15$ fm, $a=0.7$ fm and supplemented 
with a spin-orbit term, with the usual Woods-Saxon derivative form, and 
parameters $V_{so}=2.5$ MeV, $r_{so}=1.15$ fm, $a_{so}=0.7$ fm in 
order to  place the $d-$wave resonances correctly. The $d-^{209}$Bi
and  $\alpha-^{209}$Bi optical potentials are taken from Refs.~\cite{Han06} 
and \cite{Barnett74}, respectively. Consistently with 
previous works, we find that these calculations tend to underestimate the 
elastic data. We 
found that, by removing the surface part of the $d-^{209}$Bi imaginary 
potential, a good description of the experimental elastic angular 
distributions is achieved. This is shown in Fig.~\ref{6liela} by solid 
lines. For comparison, we have included the optical model calculation 
using the potential of Cook \cite{Cook82} (dashed lines).  We note that this reduction of the imaginary potential is consistent with the conclusions  of Ref.~\cite{Wat12}, which points toward an effective suppression of the deuteron breakup in $^{6}$Li scattering, compared to the free deuteron scattering.

We discuss now the inclusive breakup cross sections ($^{6}$Li,$\alpha$X). 
The EBU contribution was obtained from the CDCC calculations discussed above. 
For the NEB calculations, we used Eq.~(\ref{eq:iav}), both in the  ZR and 
FR-DWBA approximations. We adopt the same optical potential of 
$\alpha$/$d$+$^{209}$Bi as used  in the CDCC calculations. For simplicity, 
the deuteron and target spins are ignored  
(note that, in the CDCC calculations, the inclusion of the deuteron spin 
is important to place correctly the $\ell=2$ resonances). The distorted 
waves for the incoming channel are calculated with the optical potential 
of Cook quoted above. 

\begin{figure}[tb]
\begin{center}
 {\centering \resizebox*{0.8\columnwidth}{!}{\includegraphics{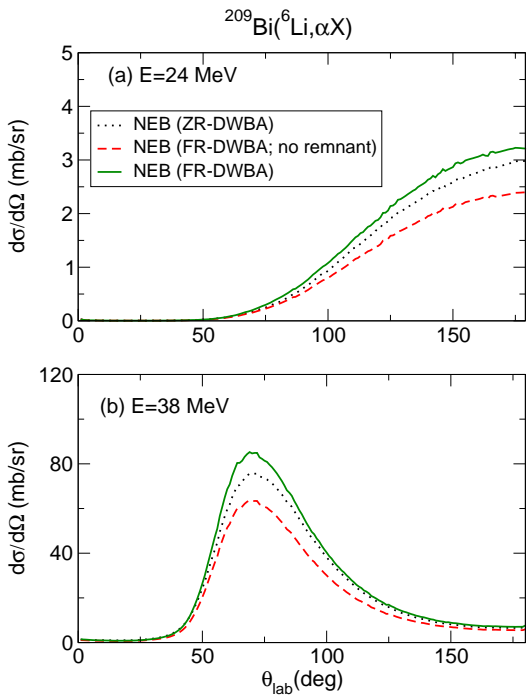}} \par}
\caption{\label{li6bi_com}(Color online) Angular 
distribution of $\alpha$ particles produced by non-elastic breakup (NEB) in the reaction 
$^6$Li+$^{209}$Bi at the incident energies of (a) 24 MeV and (b) 38 MeV. The dotted, dashed and solid 
lines are the ZR-DWBA, FR-DWBA without 
remnant term and full FR-DWBA calculations, respectively.}
\end{center}
\end{figure}

In Fig.~\ref{li6bi_dsdw}, we compare the  calculated and experimental 
angular distributions of $\alpha$ particles, for several incident energies 
of $^{6}$Li. The dotted and dashed lines are the EBU (CDCC) and  NEB (FR-DWBA)  results.
Except for the lowest energies, the NEB is found to account for most of 
the inclusive breakup cross section, in agreement with previous findings 
\cite{Kumawat10,Sig03}. The summed EBU + NEB cross sections 
(thick solid lines) reproduce fairly well the shape and magnitude of the 
data, both above and below the barrier. These results give confidence on 
the possibility of extending the formulation of the IAV theory to situations 
in which the unobserved particle is a composite system. 
  
At the most forward angles (where the $\alpha$ yield is nevertheless small) 
the EBU is found to be larger than the NEB part. Using a semi-classical 
picture, this can be understood by noting that  these small angles will 
correspond to distant trajectories. However, according to Eq.~(\ref{eq:iav}), 
the NEB is only effective for distances within the range of the 
deuteron-target imaginary potential and hence it will be very small 
for these distant trajectories. 
 It is worth noting, however, that the separation between EBU and NEB 
 parts in the ($^6$Li,$\alpha$X) case is less clear than in the ($d$,$p$X) 
 case. In the present model, the NEB is associated with the absorption 
 due to the $d$+target imaginary potential. If an empirical deuteron-target 
 potential is used, part of this absorption will be due to the breakup of 
 the deuteron into $p$+$n$. However, in a more realistic description of 
 $^6$Li in terms of $\alpha$+$p$+$n$, the breakup of $^6$Li into 
 $\alpha$+$p$+$n$ (leaving the target in the ground state) would 
 actually correspond to elastic breakup. Despite this ambiguity, 
 we believe that the sum of the two contributions, that is, the TBU cross section, can be  reasonably well estimated by the 
 present model, as supported by the comparison with the data.  

 We study now the validity of the ZR approximation in the present reaction. This is shown in Fig.~\ref{li6bi_com}, were we show the angular distribution of $\alpha$ particles produced by NEB, calculated with different  DWBA approximations, and at two different energies, one below (24 MeV) and one above (38 MeV) the barrier. 
The dotted, dashed and solid lines are the ZR-DWBA, FR-DWBA without remnant term and full FR-DWBA results, respectively. We see that the ZR-DWBA calculations underestimate
systematically the FR-DWBA results by about $\sim10-20\%$ and hence the 
validity of the ZR approximation is more questionable than in the deuteron case. Further, 
we  find that the no-remnant FR-DWBA calculation underestimates  the full FR-DWBA result  by about 
$\sim30-40\%$, indicating that the effect of the remnant term is much more important
than in the deuteron case, owning to the  strong Coulomb interaction and the 
difference of the geometry, $\vecr_{bA}$ and $\vecr_{b}$, 
caused by the valence particle.

\begin{figure}[tb]
\begin{center}
 {\centering \resizebox*{0.95\columnwidth}{!}{\includegraphics{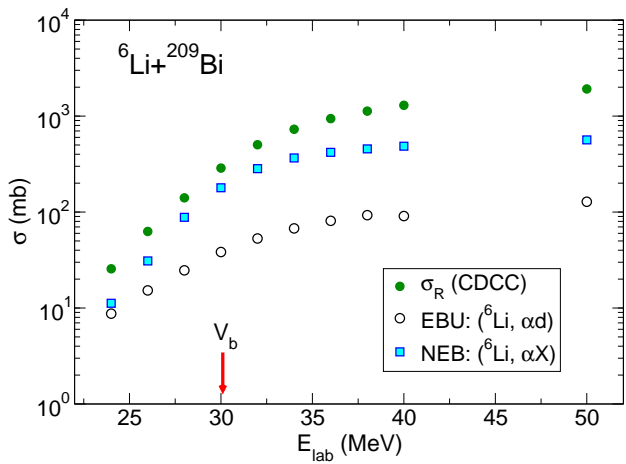}} \par}
\caption{\label{li6bi_sigedep}(Color online) Integrated cross sections for the reaction $^{6}$Li on $^{209}$Bi as a function of the incident laboratory energy. The open circles and the squares are the EBU (CDCC) and NEB (FR-DWBA) contributions to the $\alpha$ inclusive cross section. The solid circles are the reaction cross sections, obtained from the CDCC calculations. The arrow indicates the nominal position of the Coulomb barrier.}
\end{center}
\end{figure}

Finally, we study the incident energy dependence of the total $\alpha$ yield. This is shown in 
Fig.~\ref{li6bi_sigedep}. The squares and the open circles correspond, respectively, to the NEB (FR-DWBA) and EBU (CDCC) contributions 
to the $\alpha$ production cross section. 
At energies above the nominal Coulomb barrier (indicated by the arrow) 
the NEB largely dominates the inclusive breakup. Below the Coulomb barrier, 
both contributions become comparable. This can be again explained in 
classical terms, by noting that, at these small energies, the distance of 
closest approach will be relatively large, due to the presence of the 
Coulomb barrier and, therefore, the imaginary part of the $d$+target 
potential (which is responsible for the NEB part) will have little effect. We have included in the same plot 
the total reaction cross sections, as extracted from the CDCC calculations, which are found to be very close to the values calculated with the Cook 
optical potential (not shown).  It is seen that, at energies below the 
Coulomb barrier, the reaction cross section is almost exhausted by the 
($^{6}$Li,$\alpha$X) TBU cross section, whereas at energies 
above the Coulomb barrier other processes beyond the 
breakup seem to be present  (e.g.~pure target excitation, $\alpha$ absorption,  complete fusion, etc). A more detailed analysis of these processes is under study and will be presented elsewhere.

\section{Summary and conclusions \label{sec:sum}}
In summary, we have addressed the problem of the calculation of inclusive 
breakup in reactions induced by weakly-bound projectiles. For that purpose, 
we have revisited the model proposed by Ichimura, Austern and Vincent in the 
eighties \cite{Ich85,Aus87}. We have presented an alternative derivation of 
the non-elastic breakup (NEB) formula, based on a direct application of the
 coupled-channels optical theorem, which provides a transparent interpretation 
 of the NEB as the part of the flux that leaves the elastic breakup 
 channels to more complicated configurations of the $x$+$A$ system. 

Using the DWBA version of this formula,   for the NEB, and the CDCC framework, 
for the EBU part,  we have performed calculations for deuteron and $^{6}$Li 
reactions on several targets, and at different energies, finding a satisfactory 
agreement with the available inclusive breakup data in all the cases considered. 
These calculations show that, except for the particles emitted at small 
angles, most of the inclusive breakup corresponds to NEB. 
We have also tested the validity of the zero-range  approximation and the effect of the 
remnant term in the NEB calculation, by comparing with exact finite-range  DWBA calculations. 
For the studied deuteron reactions, the effect of the remnant term 
has been found to be very small and the zero-range calculation gives a result very close to the full finite-range calculation. On the other hand, 
for the $^{6}$Li+$^{209}$Bi reaction, finite-range effects become important and should be therefore considered for a correct interpretation of experimental data.

The good agreement between the calculated inclusive cross sections and 
the data suggests that this approach could be also useful to estimate the 
amount of incomplete fusion  (ICF) from the inclusive breakup.  This 
problem is of interest, for instance, in surrogate nuclear reactions 
studies \cite{Esc12}. The separation of complete from ICF  has been 
also pointed out to be  essential for the extraction of meaningful 
conclusions regarding the effect of breakup  on fusion \cite{Das04}.  
To answer these problems in a quantitatively way, one needs to extend 
the present model in order to disentangle the ICF part from other NEB 
channels, such as transfer to bound states or target excitation.

 An interesting question that arises is how these results depend on the 
 incident energy, the mass of the target and the separation energy of 
 the projectile. Further calculations are in progress in order to answer 
 these questions. In particular, in the case of the scattering of very 
 weakly-bound projectiles with heavy targets, there are evidences that 
 the EBU component can be dominant \cite{Dip12,Fer13}. A proper understanding 
 of these reactions, however, may require going beyond the DWBA 
 approximation adopted here for the NEB calculations. For that, the 
 three-body model of Austern {\it et al.} may provide an adequate framework 
 and, hence, its implementation is currently under investigation.

\begin{acknowledgments}
We are grateful to Prof.~H.~Wolter for a critical reading of the manuscript. 
This work has been partially supported by the Spanish Ministerio de Econom\'ia y Competitividad, under grant  FIS2013-41994-P,  
by the Spanish Consolider-Ingenio 2010 Programme CPAN
(CSD2007-00042)  and by Junta de  Andaluc\'ia (FQM160, P07-FQM-02894).
J.L.~is partially supported by a grant funded by the China 
Scholarship Council. 
\end{acknowledgments}


\appendix

\section{Nonelastic breakup formula in the zero-range approximation \label{sec:zr}}
In DWBA, the source term of Eq.~(\ref{eq:pz})  can be written as
\begin{equation}
\label{eq:rho_dwba}
\rho(\vec{k}_b,\vec{r}_x)=(\chi_b^{(-)}|V_\mathrm{post}|\chi_a^{(+)}\phi_a  \rangle ,
\end{equation}
with  $V_\mathrm{post}\equiv V_{bx}+U_{bA}-U_{bB}$ and where we have omitted the dependence  on $\vec{k}_a$ for simplicity of the notation. The symbol 
$(|\rangle$ means that the integration is to be taken over all
the coordinates except the $x$-channel coordinate $\vec{r}_x$. 
 If the remnant term $U_{bA}-U_{bB}$ is small, $\phi_a(\vec{r}_{bx})$ 
corresponds to an $s$-wave and $V_{bx}$ 
is short-ranged, the integral is dominated by the values 
$r_{bx} \approx 0$ and can be evaluated in the zero-range approximation, i.e.,
\begin{equation}
V_\mathrm{post} \phi_a(\vec{r}_{bx}) \simeq V_{bx}(r_{bx}) \phi_a(\vec{r}_{bx}) \simeq D_0 \delta(\vec{r}_{bx}) ,
\end{equation}
where $D_0$ is the {\it zero-range constant}. Using this approximation in (\ref{eq:rho_dwba}), and including the so-called 
{\it finite-range correction} (see, for instance, Sec.~6.14 of Ref.~\cite{satchler83}) the source term results
\begin{equation}
\rho(\vec{k}_b,\vec{r}_x)=D_0 \chi_b^{(-)*}(\vec{k}_b,c\vec{r}_x)\chi_a^{(+)}(\vec{k}_a,\vec{r}_x)\Lambda(r_x) ,
\end{equation}
where $c=m_A/(m_A+m_x)$ and  $\Lambda(r_x)$ is the {\it finite-range correction factor}. 

Ignoring the internal spins of the colliding particles, the distorted waves can be expanded as
\begin{equation}
\chi^{(+)}(\vec{k},\vec{r})= \frac{4 \pi}{k r} \sum_{l m} i^{l} R_{l}(r) Y_{l}^{m}(\hat{r}) Y_{l}^{m*}(\hat{k}).  
\end{equation}
For charged particles, the radial part is here assumed to include the Coulomb phase, $e^{i \sigma_l}$, where $\sigma_l$ are the Coulomb phase shifts.   

Following \cite{Kas82}, the source term is expanded in  spherical harmonics as 
\begin{align}
\rho(\vec{k}_b,\vec{r}_x) & =\frac{16\pi^{2}}{k_ak_b}\sum_{l_xm_x}Y_{l_x}^{m_x}(\hat{r}_x) \nonumber \\ 
& \times \sum_{l_a}\sum_{l_b}
\rho_{l_x}^{l_al_b}(r_x)\mathfrak{Y}_
{l_xm_x}^{l_al_b}(\hat{k}_a,\hat{k}_b)
\end{align}
with 
\begin{align}
\label{eq:rho_expand}
\rho_{l_x}^{l_al_b}(r_x)& =\frac{D_0}{cr_x^2}i^{l_a+l_b}(-1)^{l_b}\Big[
\frac{(2l_a+1)(2l_b+1)}{4\pi(2l_x+1)}\Big]^{1/2} \nonumber \\
& \times \langle l_al_b00|l_x0\rangle 
R_{l_a}(r_x)R_{l_b}(cr_x)\Lambda(r_x)
\end{align}
and 
\begin{equation}
\mathfrak{Y}_{l_xm_x}^{l_al_b}(\hat{k}_a,\hat{k}_b) 
=\sum_{m_am_b}\langle l_al_bm_am_b|l_xm_x\rangle Y_{l_b}^{m_b*}(\hat{k}_b)
Y_{l_a}^{m_a*}(\hat{k}_a)
\end{equation}
The channel wave function $\psix$ 
in Eq.~(\ref{eq:pz}) is also expanded in spherical harmonics
\begin{equation}
\psix=\frac{16\pi^2}{k_ak_b}\frac{1}{r_x} \sum_{l_xm_x} \psi_{l_xm_x}^{0}(r_x,\vec{k}_a,\vec{k}_b)Y_{l_xm_x}(\hat{r}_x)
\end{equation}
For convenience, $\psi_{l_xm_x}^{0}(r_x,\vec{k}_a,\vec{k}_b)$ is written as
\begin{equation}
\label{eq:psi_expand}
\psi_{l_xm_x}^{0}(r_x,\vec{k}_a,\vec{k}_b)=
\sum_{l_a}\sum_{l_b}\mathfrak{R}_{l_x}^{l_al_b}(r_x)
\mathfrak{Y}_{l_xm_x}^{l_al_b}(\hat{k}_a,\hat{k}_b) .
\end{equation}
Inserting the expansions (\ref{eq:rho_expand}) and (\ref{eq:psi_expand}) into 
the inhomogeneous equation Eq.~(\ref{eq:pz}), one gets
\begin{equation}
\Bigg\{\frac{\hbar^2}{2\mu_x}\Big[\frac{\mathrm{d}^2}
{\mathrm{d}r_x^2}-\frac{l_x(l_x+1)}{r_x^2}
\Big]-U_x+E_x\Bigg\}\mathfrak{R}_{l_x}^{l_al_b}(r_x)=r_x \, \rho_{l_x}^{l_al_b}(r_x)
\end{equation}
For $E_x>0$ (unbound $x$-$A$ states), this equation is to be solved with outgoing boundary conditions
\begin{equation}
\mathfrak{R}_{l_x}^{l_al_b}(r_x) \rightarrow - S^{l_a,l_b}_{l_x} H^{(+)}_{l_x}(k_x r_x)
\end{equation}
where $H^{(+)}_{l_x}(k_x r_x)$ is a Coulomb outgoing wave and the coefficients $S_{l_x}^{l_a,l_b}$ are the S-matrix elements. 

Finally, the double differential cross section of nonelastic breakup 
with zero-range approximation results
\begin{equation}
\Big[\frac{\mathrm{d}^2\sigma}{\mathrm{d}\Omega_b\mathrm{d}E_b}\Big]
_{\mathrm{post}}^{\mathrm{NEB}}=\frac{64\pi\mu_a\mu_b}
{\hbar^4k_a^3k_b}\sum_{l_xm_x}\mathfrak{I}_{l_xm_x}
(\vec{k}_a,\vec{k}_b)
\end{equation}
where 
\begin{widetext}
\begin{equation}
\mathfrak{I}_{l_xm_x}(\vec{k}_a,\vec{k}_b)=\int \mathrm{d}r W_x(r_x)\bigg\lvert
\sum_{l_a}\sum_{l_b}
\mathfrak{R}_{l_x}^{l_al_b}(r_x)\mathfrak{Y}_{l_xm_x}^{l_al_b}(\hat{k}_a,\hat{k}_b)
\bigg\rvert^2
\end{equation}
\end{widetext}

\section{Nonelastic breakup formula in the finite-range approximation \label{sec:fr}}
In the finite-range approximation, the source term (\ref{eq:rho_dwba}) 
is evaluated exactly. Because all the relevant coordinates lie on the same 
plane (see Fig.~\ref{zrcoor}), one can express any coordinate in terms of 
two independent vectors. So, for example, choosing $\vec{r}_x$ and $\vec{r}_{b}$ 
as independent vectors, one may write
\begin{equation}
\vec{r}_{bx}=q\vec{r}_x-\vec{r}_{b} \ \ \ \mathrm{and} \ \ \   \vec{r}_a=(1-pq)\vec{r}_x+p\vec{r}_{b}
\end{equation}
where $p=m_b/(m_b+m_x)$ and $q=m_A/(m_A+m_x)$. The projectile wave function, 
neglecting again its internal spin, can be expressed as 
$\phi_a(\vec{r}_{bx}) = (R_{l_{bx}}(r_{bx})/r_{bx}) Y_{l_{bx}}^{m_{bx}}(\hat{r}_{bx})$. 
Using this, and the partial wave decomposition of the distorted waves, the source term is written as
\begin{align}
\rho_b^{m_{bx}}(\vec{k}_b,\vec{r}_x) 
& = \frac{16\pi^2}{k_ak_b}\sum_{l_am_a}\sum_{l_bm_b}i^{l_a+l_b}(-1)^{l_b}
Y_{l_b}^{m_b*}(\hat{k}_b) 
\nonumber \\
& \times Y_{l_a}^{m_a*}(\hat{k}_a) \int \mathrm{d}
\vec{r_{b}} V_\mathrm{post} \frac{R_{l_b}(r_b)}{r_b}Y_{l_b}^{m_b}(\hat{r}_b)
\nonumber \\
&  \times \frac{R_{l_a}(r_a)}{r_a} Y_{l_a}^{m_a}(\hat{r}_a)
\frac{R_{l_{bx}}(r_{bx})}{r_{bx}}Y_{l_{bx}}^{m_{bx}}(\hat{r}_{bx})
\end{align}
To calculate this, we  transform the spherical harmonics 
$Y_{l_a}^{m_a}(\hat{r}_a)$ and $Y_{l_{bx}}^{m_{bx}}(\hat{r}_{bx})$ into 
linear combinations of the spherical harmonics $Y_{l_b}^{m_b}(\hat{r}_b)$ 
and $Y_{l_x}^{m_x}(\hat{r}_x)$. This is done by means of the Moshinsky 
solid-harmonic expansion \cite{Moshinsky59}
\begin{widetext}
\begin{align}
Y_{l_{bx}}^{m_{bx}}(\hat{r}_{bx}) & =  \sqrt{4\pi}\sum_{n=0}^{l_{bx}}\sum_
{\lambda=-n}^nc(l_{bx},n)
 \frac{(qr_x)^{l_{bx}-n}(-r_{b})^n}{r_{bx}^{l_{bx}}}
Y_{l_{bx}-n}^{m_{bx}-\lambda}(\hat{r}_x)Y_n^{\lambda}(\hat{r}_{b}) 
\langle l_{bx}-n,n,m_{bx}-\lambda,\lambda|l_{bx},m_{bx}\rangle ,
\end{align}
%
\begin{equation}
\begin{split}
Y_{l_a}^{m_a}(\hat{r}_a) & =\sqrt{4\pi}\sum_{u=0}^{l_a}\sum_{\nu=-u}^u
c(l_a,u) 
\frac{(pr_{b})^{l_a-u}(1-pq)^u(r_{x})^u}{r_a^{l_a}}
Y_{l_a-u}^{m_a-\nu}(\hat{r}_b) 
Y_u^{\nu}(\hat{r}_{x})\langle l_a-u,u,m_a-\nu,\nu|l_a,m_a\rangle ,
\end{split}
\end{equation}
\end{widetext}
where
\begin{equation}
c(x,y)=\Bigg(\frac{(2x+1)!}{(2y+1)!(2(x-y)+1)!}\Bigg)^{1/2}
\end{equation}
Because the interaction $V_\mathrm{post}$ is an scalar, we can perform 
the Legendre expansion
\begin{equation}
V_\mathrm{post} \frac{R_{l_a}(r_a)}{(r_a)^{l_a+1}}\frac{R_{l_{bx}}(r_{bx})}{(r_{bx})^
{l_{bx}+1}}=
\sum_{T=0}^{T_{max}}(2T+1)\mathbf{q}_{l_a,l_{bx}}^{T}(r_{b},r_x)P_T(z)
\end{equation}
We note that, even if a finite-range treatment is made, in reactions of light projectiles on heavy targets (e.g., deuteron scattering on heavy targets),  
the difference  $U_{bA}-U_{bB}$, known as {\it remnant term}, can be neglected, and thus $V_\mathrm{post}\simeq V_{bx}$. 
The limit $T_{max}$ is chosen large enough to generate all the couplings 
for partial waves to be used. Here, the argument $z$ in the Legendre polynomials 
$P_T(z)$ is the cosine of the angle between $\vec{r}_b$ and $\vec{r}_x$. The radial kernels are explicitly given by 
\begin{equation}
\mathbf{q}_{l_a,l_{bx}}^{T}(r_{b},r_x)  =\frac{1}{2}\int_{-1}^1
V_\mathrm{post} \frac{R_{l_a}(r_a)}{(r_a)^{l_a+1}}\frac{R_{l_{bx}}(r_{bx})}{(r_{bx})^
{l_{bx}+1}}  P_T(z)\mathrm{d}z .
\end{equation}

Finally, the source term results  
\begin{align}
\rho_b^{m_{bx}}(\vec{k}_b,\vec{r}_x)& =\frac{16\pi^2}{k_ak_b}\sum_{l_xm_x} Y_{l_x}^{m_x} (\hat{r}_x) \nonumber \\
&\times
\sum_{l_al_b}\sum_l  \mathfrak{Y}_{l_al_b}^{ll_xm_x{m_{bx}}} (\hat{k}_a,\hat{k}_b) 
\rho^{l_al_b}_{ll_x} (r_x) ,
\end{align}
with
\begin{align}
\mathfrak{Y}_{l_al_b}^{ll_xm_x{m_{bx}}} (\hat{k}_a,\hat{k}_b)& =\sum_{m_am_b}
Y_{l_a}^{m_a*}(\hat{k}_a)Y_{l_b}^
{m_b*}(\hat{k}_b) \nonumber \\
&  \times \langle l_al_{bx}m_am_{bx}|lm_l\rangle \langle l l_b m_l m_b | l_x m_x \rangle ,
\end{align}
and
\begin{align}
\rho^{l_al_b}_{ll_x} (r_x) 
& =\sum_{nu}\sum_{\Lambda_a
\Lambda_b}\sum_T i^{l_a+l_b}(-1)^{l_b+l+n+\Lambda_b-\Lambda_a}p^{l_a-u} \nonumber \\
&  \times (qr_x)^{l_{bx}-n}(r_x)^u(1-pq)^u
\widehat{l_a-u} \widehat{l_{bx}-n}\hat{n}\hat{u} \nonumber \\
&  \times \hat{l}_{bx}
\hat{\Lambda_a}\hat{\Lambda_b}\hat{l_a}\hat{l_b}\hat{T}/\hat{l}/\hat{l_x}
c(l_{bx},n)c(l_a,u)  \nonumber \\
& \times \langle u,l_{bx}-n,00|\Lambda_b0 \rangle 
\langle l_a-u,n,0,0|\Lambda_a,0 \rangle \nonumber \\
&  \times \langle \Lambda_b,T,0,0|l_{x},0\rangle\langle \Lambda_a,l_{b},0,0|T,0\rangle 
 (2l+1) \nonumber \\
&\times \left\{\!\begin{array}{ccc}
l_{bx} & l & l_a \\
n  & \Lambda_a & l_a-u \nonumber \\
l_{bx}-n &  \Lambda_b & u 
\end{array}\!\right\} W(l_x,\Lambda_b,l_{b},\Lambda_a;T,l) \nonumber \\
& \times   
\int \mathrm{d}r_{b}
R_{l_{b}}(r_{b})
(r_{b})^{l_a-u+n+1}
\mathbf{q}_{l_a,l_{bx}}^{T}(r_{b},r_x) 
\end{align}

As in the zero-range case, $\psi_x^0(\vec{k}_b,\vecr_x)$ can be expanded as 
\begin{align}
\psi_x^0(\vec{k}_b,\vecr_x) & =
\frac{16\pi^2}{k_ak_b}r_x^{-1}\sum_{l_{x}m_{x}}
Y_{l_{x}}^{m_{x}}(\hat{r}_x)
\nonumber \\
& \times \sum_{l_al_b}\sum_l 
\mathfrak{R}^{l_al_b}_{ll_x}(r_x)
\mathfrak{Y}_{l_al_b}^{ll_xm_x{m_{bx}}}(\hat{k}_a,\hat{k}_b)
\end{align}
where the radial coefficients, $\mathfrak{R}^{l_al_b}_{ll_x}(r_x)$, are  solutions of the inhomogeneous equation 
\begin{equation}
\Bigg\{\frac{\hbar^2}{2\mu_x}\Big[\frac{\mathrm{d}^2}
{\mathrm{d}r_x^2}-\frac{l_{x}(l_{x}+1)}{r_x^2}
\Big]-U_x+E_x\Bigg\}\mathfrak{R}^{l_al_b}_{ll_{x}}(r_x)=r_x\rho^{l_al_b}_{ll_{x}}(r_x)
\end{equation}
The boundary condition is the same as in the zero-range case.

 Finally, the double 
differential cross section with finite range post-form DWBA can be written as 
\begin{equation}
\Big[\frac{\mathrm{d}^2\sigma}{\mathrm{d}\Omega_b\mathrm{d}E_b}\Big]_
{\mathrm{post}}^{\mathrm{NEB}}=\frac{64\pi\mu_a\mu_b}
{\hbar^4k_a^3k_b}\frac{1}{2l_{bx}+1}\sum_{l_{x}m_{x}}\mathfrak{I}_{l_{x}m_{x}}^{m_{bx}}
(\vec{k}_a,\vec{k}_b)
\end{equation}
with
\begin{widetext}
\begin{equation}
\mathfrak{I}_{l_{x}m_{x}}^{m_{bx}}(\vec{k}_a,\vec{k}_b)=\int \mathrm{d}r_x W_x(r_x)
\bigg\lvert\sum_{l_al_bl} \mathfrak{R}^{l_al_b}_{ll_{x}}(r_x)
\mathfrak{Y}_{l_al_b}^{ll_xm_x{m_{bx}}}(\hat{k}_a,\hat{k}_b)\bigg\rvert^2
\end{equation}
\end{widetext}


\bibliography{inclusive_prc.bib}
\end{document}